\documentclass{ws-p9-75x6-50}

\def\laq{~\raise 0.4ex\hbox{$<$}\kern -0.8em\lower 0.62
ex\hbox{$\sim$}~}
\def\gaq{~\raise 0.4ex\hbox{$>$}\kern -0.7em\lower 0.62
ex\hbox{$\sim$}~}

\begin{document}


\begin{flushright}
BA-TH/00-397\\
gr-qc/0009098
\end{flushright}

\vspace*{0.8truein}

{\Large\bf\centering\ignorespaces
Production and Detection of Relic Dilatons \\ 
\bigskip
in String Cosmology
\vskip2.5pt}
{\dimen0=-\prevdepth \advance\dimen0 by23pt
\nointerlineskip \rm\centering
\vrule height\dimen0 width0pt\relax\ignorespaces

\vspace{0.8 cm}
M. Gasperini
\par}
\vspace{0.5 cm}
{\small\it\centering\ignorespaces
Dipartimento di Fisica, Universit\`a di Bari,\\
Via G.  Amendola 173, 70126 Bari, Italy \\
\vspace{0.2 cm}
and\\
\vspace{0.2 cm}
Istituto Nazionale di Fisica Nucleare,  Sezione di Bari, Bari, Italy\\
\par}

\par
\bgroup
\leftskip=0.10753\textwidth \rightskip\leftskip
\dimen0=-\prevdepth \advance\dimen0 by17.5pt \nointerlineskip
\small\vrule width 0pt height\dimen0 \relax

\vspace*{0.6truein}

\centerline{\bf Abstract}

\noindent
This paper summarizes the contribution presented at the {\sl IX Marcel
Grossmann Meeting} (Rome, July 2000). It is stressed, in particular, that
a non-relativistic background of ultra-light dilatons, produced in the
context of string cosmology, could represent today a significant
fraction of cold dark matter. If the dilaton mass lies within the
resonant band of present gravitational antennas, a stochastic dilaton
background with a nearly critical density could be visible, in
principle, already at the  level of sensitivity of the detectors
in operation and presently under construction.

\vspace{0.8cm}
\begin{center}
------------------------------  

\vspace{0.8cm}
Contributed paper to the {\bf  IX Marcel
Grossmann Meeting} (Rome, July 2000)\\ 
To appear in the Proceedings (World Scientific, Singapore)
\end{center}

\thispagestyle{plain}
\par\egroup

\vfill


\newpage
\thispagestyle{empty}
\vbox{}
\newpage

\title{Production and Detection of Relic Dilatons \\ in String Cosmology}

\author{M. Gasperini}

\address{Dipartimento di Fisica, Universit\`a di Bari, and INFN, Sezione
di Bari, \\Via G. Amendola 173, 70126 Bari, Italy
\\E-mail: gasperini@ba.infn.it}


\maketitle

\abstracts{\centerline{Preprint No. BA-TH/00-397;  ~~~~~~~E-print
Archives: gr-qc/0009098}}


The main purpose of this paper is to point out that the relic dilatons
produced in the context of the pre-big bang scenario \cite{1} could
provide a significant contribution to the present fraction of dark
matter density and, if light enough, could be detectable even by
present gravitational antennas \cite{2}, provided their coupling to
macroscopic bodies is not too far from the present experimental upper
limits. 

The spectrum of the relic dilatons, produced from the vacuum
through the parametric amplification of quantum fluctuations, differs
from the graviton spectrum because of the mass of the dilaton, and
contains in general  three branches \cite{3}, corresponding to 
\begin{itemize}
\item{}
relativistic modes, with proper momentum $p=k/a > m$;  
\item{}
non-relativistic modes, with $p_m < p<m$, which were still
relativistic at the time of horizon crossing; 
\item{}
non-relativistic
modes, with $p < p_m$, which were already non-relativistic at the time
of horizon crossing. 
\end{itemize}
The last two branches are separated by the limiting
frequency $p_m= p_1 (m/H_1)^{1/2}$ of a mode that becomes
non-relativistic, $p \simeq m$, just at the time of horizon crossing, $p
\simeq H$ (here $p_1=H_1a_1/a$ is the maximal amplified momentum
scale, and $H_1$ the curvature scale at the time $t=t_1$ of the
inflation-radiation transition). 

At the present time $t_0$, the
``minimal" dilaton spectrun can thus be written as follows \cite{3}: 
\begin{equation}
\begin{array}{rcl}
&&\\
\Omega_\chi(p,t_0) &&{} \simeq   \left(H_1\over M_p\right)^2 
\Omega_\gamma (t) \left(p\over p_1\right)^\delta, 
~~~~~~~~~~~~~~~~ ~m< p <p_1, 
\\[4pt]
&&{}\simeq   \left(H_1\over M_p\right)^2 
\left(m^2\over H_1H_{\rm eq}\right)^{1/2} \left(p\over
p_1\right)^{\delta-1},  ~~~~~~ p_m< p <m, 
\\[4pt]
&&{}\simeq   \left(H_1\over M_p\right)^2 
\left(m\over H_{\rm eq}\right)^{1/2} \left(p\over
p_1\right)^{\delta},  ~~~~~~~~~~~~ p< p_m, 
\\&&
\end{array}
\label{1}
\end{equation}
where $M_p$ is the Planck mass, $H_{\rm eq} \sim 10^{-55}M_p$ is the
curvature scale at the epoch of matter-radiation equilibrium, and
$\Omega_\gamma (t)= (H_1/H)^2(a_1/a)^4$ is the energy density (in
critical units) of the radiation that becomes dominant at the end of
inflation (today, $\Omega_\gamma (t_0) \sim 10^{-4}$).
Finally, $\delta$ is a slope parameter depending on the kinematics of
the pre-big bang phase ($\delta =3$ for dilaton-dominated inflation
\cite{1}, while $\delta <3$ for modes crossing the horizon during the
high curvature string phase). 

Let us suppose now that the dilaton mass
is small enough to fall inside the sensitivity band of present
gravitational detectors, so that $m \ll p_1$, and assume $\delta >0$, to
avoid infrared divergences. It follows that the relativistic branch of
the spectrum is growing, and it is bounded today by the peak value 
$\Omega_\chi^{\rm rel} \sim (H_1/M_p)^2 \Omega_\gamma \laq
10^{-6}$, where we have used the fact that the final inflation scale is
controlled by the string mass scale $H_1 \simeq M_s \sim (0.1 - 0.01)
M_p$. 

The non-relativistic part of the spectrum, on the contrary, is only
constrained by the critical density bound, 
\begin {equation} 
\int^m d \ln p
\Omega_\chi^{\rm non-rel}(p) \laq 1. 
\end{equation}
Let us suppose that $\delta <1$,
so that the above integral is dominated by the peak of the spectrum at
$p=p_m$, and  leads to the constraint 
\begin {equation} 
\Omega_\chi (p_m) \simeq \left(H_1\over M_p \right)^2
\left(m\over H_{\rm eq}\right)^{1/2} \left(m\over H_1\right)^{\delta/2}
\laq 1.  
\end{equation}
If this constraint is
saturated, i.e. if the dilaton mass satisfies
\begin{equation}
m \simeq (H_{\rm eq} M_p^4 H_1^{\delta-4})^{1/(\delta+1)}, 
\label{2}
\end{equation}
then the non-relativistic branch dominates the dilaton spectrum, and
the relic dilatons might represent a significant fraction of the present
cold dark matter density. It is important to stress that, for a nearly flat
spectrum ($\delta \rightarrow 0$), this may occurr even for very low
masses, lying in the sensitivity range of gravitational antennas. If we
take, for instance, $m \sim 1$ kHz $\sim 10^{-12}$ eV, and $H_1=M_s
\sim 10^{-1}M_p$, then the condition (\ref{2}) is satified for $\delta
\simeq 11/39 \simeq 0.28$. 

It is not impossible, therefore, that the very weak (much
smaller than gravitational) coupling to the detectors of  ultra-light
dilatons \cite{4} may be compensated by a very high (almost critical)
relic energy density (much higher than for relic gravitons). If this is the
case, an analysis of the signal-to-noise ratio produced by a
non-relativistic stochastic background of scalar particles \cite{2}
shows that the detection of a dilatonic dark matter component is in
principle compatible with the expected sensitivity of the gravitational
antennas in operation and presently under construction.

\end{document}